\documentclass[fleqn,10pt]{wlscirep}

\usepackage[utf8]{inputenc}
\usepackage[T1]{fontenc}
\usepackage{bm}
\usepackage{booktabs}
\usepackage{tabularx}
\usepackage{pifont}
\usepackage{graphicx}
\usepackage{xcolor}
\usepackage{mdframed}
\usepackage{marvosym}

\newcommand{\cmark}{\ding{51}}
\newcommand{\xmark}{\ding{53}}

\definecolor{ly}{RGB}{255,255,200}

\usepackage[printwatermark]{xwatermark}
\usepackage{tikz}

\title{The Physics of Financial Networks}

\author[1,2,*]{Marco Bardoscia}
\author[2]{Paolo Barucca}
\author[3,4]{Stefano Battiston}
\author[2,5,6]{Fabio Caccioli}
\author[7,8,9]{Giulio Cimini}
\author[8,10]{Diego Garlaschelli}
\author[8]{Fabio Saracco}
\author[8]{Tiziano Squartini}
\author[11,12,9,13,\Letter]{Guido Caldarelli}
\affil[1]{Bank of England, London, UK}
\affil[2]{University College London, Department of Computer Science, London, UK}
\affil[3]{University of Zurich, Department of Banking and Finance, Zurich, Switzerland}
\affil[4]{University of Venice ``Ca' Foscari'', Department of Economics, Venice, Italy}
\affil[5]{Systemic Risk Centre, London School of Economics and Political Sciences, London, UK}
\affil[6]{London Mathematical Laboratory, London, UK}
\affil[7]{University of Rome ``Tor Vergata'', Department of Physics and INFN, Rome, Italy}
\affil[8]{IMT School for Advanced Studies, Networks Unit, Lucca, Italy}
\affil[9]{Consiglio Nazionale delle Ricerche, Institute of Complex Systems, Rome, Italy}
\affil[10]{Lorentz Institute for Theoretical Physics, University of Leiden, Leiden, The Netherlands}
\affil[11]{Department of Molecular Science and Nanosystems, University of Venice ``Ca' Foscari'', Venice, Italy}
\affil[12]{European Centre of Living Technologies, University of Venice ``Ca' Foscari'', Venice, Italy}
\affil[13]{London Institute for Mathematical Sciences, London, UK}
\affil[*]{Any views expressed are solely those of the author(s) and so cannot be taken to represent those of the Bank of England or to state
Bank of England policy.}
\affil[\Letter]{Guido.Caldarelli@unive.it}

\begin{abstract}
The field of Financial Networks is a paramount example of the novel applications of Statistical Physics that have made possible by the present data revolution. 
As the total value of the global financial market has vastly outgrown the value of the real economy,
financial institutions on this planet have created a web of interactions whose size and topology calls for a quantitative analysis by means of Complex Networks. 
Financial Networks are not only a playground for the use of basic tools of statistical physics as ensemble representation and entropy maximization; rather, their particular dynamics and evolution triggered theoretical advancements as the definition of DebtRank to measure the impact and diffusion of shocks in the whole systems. In this review we present the state of the art in this field, starting from the different definitions of financial networks (based either on loans, on assets ownership, on contracts involving several parties -- such as credit default swaps, to multiplex representation when firms are introduced in the game and a link with real economy is drawn) and then discussing the various dynamics of financial contagion as well as applications in financial network inference and validation.
We believe that this analysis is particularly timely since financial stability as well as recent innovations in climate finance, once properly analysed and understood in terms of complex network theory, can play a pivotal role in the transformation of our society towards a more sustainable world.
\end{abstract}

\begin{document}

\flushbottom

\maketitle

\thispagestyle{empty}

\section{Introduction}

Statistical physics provides a mathematical description of the relation between microscopic and macroscopic properties of physical systems composed by many parts. For this reason, in the last decades, statistical physics has emerged as a legitimate approach to investigate such a relation in social and economic systems \cite{Lazer721,pentland2014social,buchanan2008social,ball2012society,caldarelli2018physics}. In particular, the study of complex networks and their applications to economics and finance has become a  playground for the discipline  \cite{auyang1998foundations,mantegna1999introduction,barabasi2012network,RevModPhys.81.591,RevModPhys.87.925,perc2017statistical,cimini2019statistical}. Especially in the field of financial networks the application of physics to social systems has been greatly successful in terms of results and impact \cite{may2010systemic,beale2011individual,battiston2012debtrank,battiston2016complexity,cimini2015systemic,battiston2016price}. Aim of this review is to present the main research questions and results, and the future avenues of research in this field.
It is widely recognized today that modelling the financial system as a network is a precondition to understand and manage a wide range of phenomena that are not just relevant to finance professionals or scholars, but also to researchers in many other disciplines, ordinary citizens, public agencies and governments \cite{stiglitz2010risk,haldane2011systemic}.
Indeed, the implications of dysfunctions in the financial system include: ordinary citizens losing jobs and savings as in the 2009 financial crisis; public budgets going into deficit (thus shrinking also funding for scientific research) as in the 2012 sovereign crisis; excessive investments flowing into carbon-intensive economic activities, thus jeopardising the achievement of the climate mitigation objectives of the Paris Agreement.  

The discipline of financial networks has filled an important scientific gap, by showing that many important phenomena in the financial system emerge because of the indirect interaction of financial actors. 
This means, for example, that if the price of a certain asset plummets, not only those actors who have invested in that asset are affected, but also those who have invested in the obligations of the former actors. Because of the intricate chains of contracts, and the feedback mechanisms, the resulting effects can be much larger than the initial shocks.
As in other domains of complex systems, the emergence of system-level instabilities can only be understood from the interplay of the network structure (e.g.\ closed chains) and key properties of links and nodes (e.g.\ about risk propagation and financial leverage)\cite{bardoscia2017pathways, roukny2018interconnectedness}. 
%
Traditional economic models have described the financial system either as an aggregate entity or as a collection of actors in isolation, failing to provide an appropriate description of these mechanisms and their implications for society \cite{colander2009financial}.

The discipline of financial networks, has been very interdisciplinary from the start, positioned at the frontier of graph theory, statistical physics of networks, financial economics. 
The units of observation (e.g.\ financial actors with their balance sheets and their contracts) belong by their nature to the domain of economics and finance, as do areas of applications such as the analysis of corporate influence and systemic risk.   
Once the initial questions\cite{kogut2001small,eisenberg2001systemic,allen2000financial} have been analysed in terms of network theory, new methods and insights have emerged, in particular in the statistical physics community, that have enriched the initial set of questions: when does the network amplify or absorb risk? How does the network structure interplay with the feature of the risk propagation  process? How does the network structure and the expectations of the financial actors co-evolve?  

The discipline of financial networks has delivered statistical tools and analytical models to characterise financial risk beyond the traditional boundaries of economics by accounting for the complexity and the interconnectedness of the financial system. The key contributions to this endeavour are demonstrated by the adoption of concepts and metrics by practitioners and policy makers in the financial sector \cite{henry2013macro,abad2016shedding,churm2019four} and by scholars in the economics profession \cite{allen2009networks,elliott2014financial,acemoglu2015systemic}. 

\begin{figure}
\centering
\includegraphics[width=0.5\columnwidth]{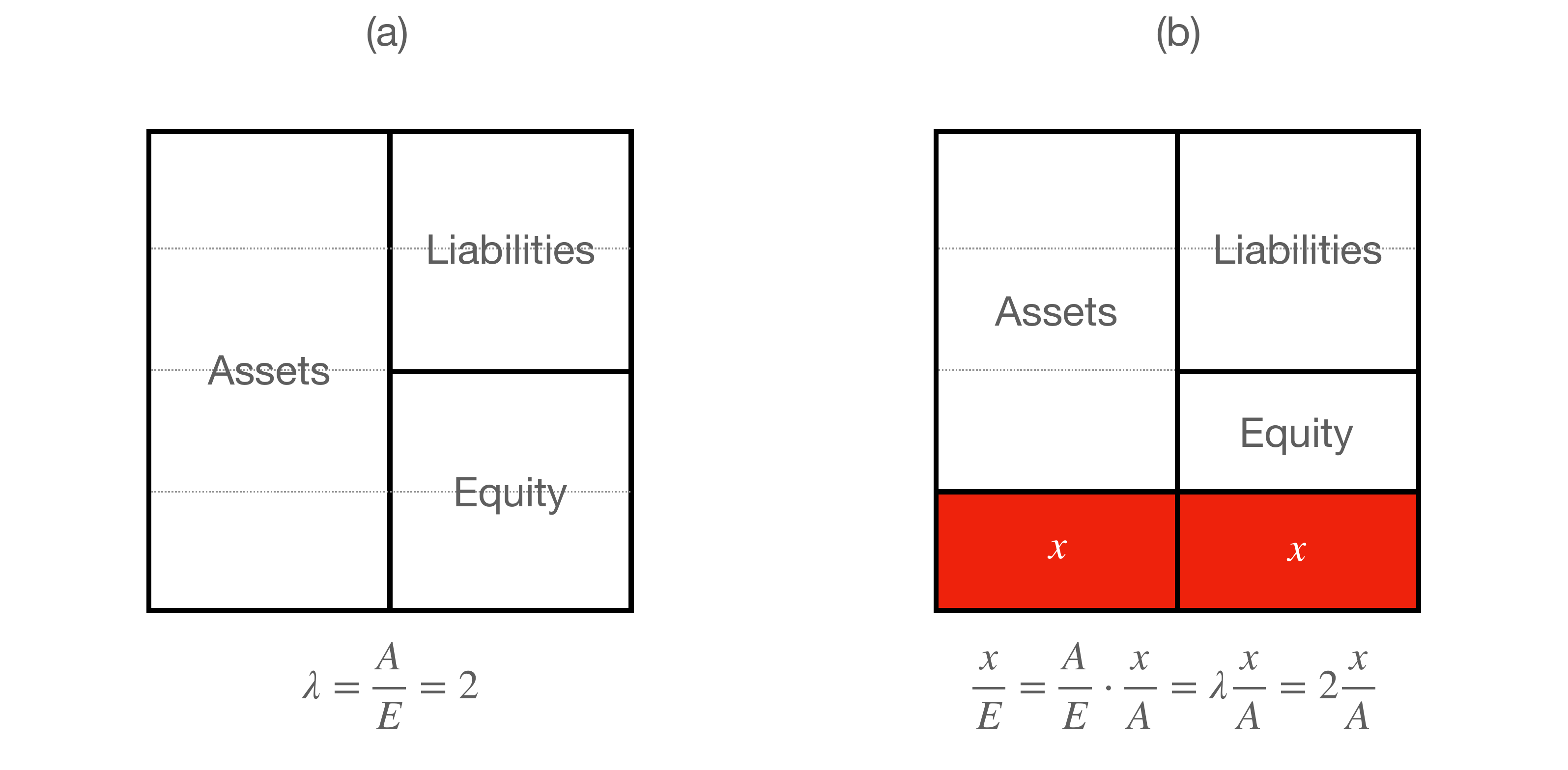}
\caption{Stylised balance sheet of an investor with leverage $\lambda = 2$ (a). Shocks to assets translate into equity shocks amplified by a factor $\lambda$ (b).}
\label{fig:leverage}
\end{figure}

\begin{mdframed}[backgroundcolor=ly]
\medskip

\section*{Box 1: Leverage}

Investors are said to be leveraged when they borrow money to invest.  
For instance, we use leverage when we get a mortgage to buy a house. 
If we put a capital of £ 40 000 as a downpayment and we borrow £ 160 000 to buy a house worth £ 200 000, then our leverage is equal to 5: the value of our assets (the house) divided by our capital.
Leverage is related to risk, because it amplifies our gains and losses. 
If the value of the house increases to £ 220 000, we could sell it, pay back our debt (let us assume for simplicity there is no interest rate), and we would have gained £ 20 000.
We see then that an increase of 10\% in the value of the house translates into an increase of 50\% of our initial capital. The same is however true if the house is devalued: a devaluation of 10\% would lead to a 50\% reduction of our initial capital (from £ 40 000 to £ 20 000). More in general, if our leverage is equal to $\lambda$, a 1\% change in the value of the house translates into a $\lambda$ \% change of our capital. 
The same applies to all leveraged investors. 
Leverage $\lambda$ is defined in general for any investor or institution as the ratio between assets and equity. 
In Figure \ref{fig:leverage} we show the stylized representation of a balance sheet of an investor with leverage equal to two. 
When the assets are devalued by $25\%$ 
the equity lost is $50\%$, equal to the asset devaluation multiplied by leverage: The higher leverage, the higher the amplification of losses, the higher the risk of the investor.
So far we have considered an isolated investor, but the concept of leverage as an amplifier of losses can be generalized to the context of a network of interconnected balance sheets. For instance, when banks lend money to each other, the interbank assets of a bank correspond to the interbank liabilities of other banks. When a bank is under stress, the value of the interbank assets associated with its liabilities are devalued, which puts its creditors under stress, and so on. It can be shown that the propagation of shocks within the network is governed by a matrix, called matrix of interbank leverage, whose leading eigenvalue determines the level of endogenous amplification of exogenous shocks~\cite{bardoscia2017pathways}.
\end{mdframed}

\section{Network structure}
\subsection{The financial system as a network}
The financial system consists of financial actors (e.g.\ institutions such as banks, or pension funds, but also small ``fintech'' companies or households), markets (e.g.\ the stock, or the bond market), contracts (e.g.\ the ownership in a stock, or a loan between two banks, or from a bank to a firm, or from a bank to a household), and regulatory bodies (e.g.\ financial supervisors and central banks). A network is thus a natural description of the financial system. Nodes represent financial actors and links may represents contracts or other types of relations (e.g.\ two actors investing in the same asset). 
There are processes occurring on the financial network, where the properties of nodes change but the link remain the same, such as the flow of revenues from economic activities to the owners of the corresponding securities, or the propagation of losses. Furthermore, in a financial network the relationships change over time. As a result, a complete description of the financial system requires a temporal multiplex network, where each layer is associated with a specific type of relationship (e.g.\ interbank loans of a given maturity). The emergent macroscopic properties of the network are then important to understand questions of general interest such as the conditions for having financial stability or a smooth transition to a low-carbon economy. However, the structure of financial networks evolves also as a result of actors trying to anticipate the future of the network itself in competition with other nodes. This feedback loop makes the investigation of the emergent properties of financial networks a fascinating scientific challenge that differs fundamentally from questions pertaining networks in the natural science domains. 

In this review we proceed in the following order. We start with the characterization of the structure of static financial networks, single and multiplex. We then review the most studied processes taking place on financial networks, focusing on financial contagion, first along bilateral links in unipartite networks, and then through common neighbouring nodes (overlapping portfolios) on bi-partite networks. Further, we cover the problem of estimating the structure of the financial networks from partial information. More in general, we review the stream of works on constructing ensembles of financial networks that satisfy certain properties on average or in distribution. 

In order to describe both the structure and the microscopic processes taking place on financial networks, we will have to introduce some specific notions and jargon. However, we will try to map them in terms of nodes, links and physical analogs in order to highlight similarities and differences with other related to other areas of physics, such as contact processes\cite{castellano2006nonmean}, diffusion theory\cite{masuda2017random}, and epidemics\cite{pastorsatorras2015epidemic}.

\subsection{Single-layer networks}

While economic networks comprise several types of relations, such as credit lending or supply of goods and services, the network of ownership best reflects the relations of power\cite{weber1978economy,hill2015research} among economic and financial actors. 
Through chains of ownership, shareholders have a means to influence, intentionally or not, the activities of firms owned directly and indirectly. Thus, one stream of work has investigated the structure of ownership networks and its implications.
Ownership networks display small world properties \cite{kogut2001small,corrado2006small,fichtner2017hidden}, scale-free network properties appear in stock markets across countries \cite{garlaschelliea2005scale-free, glattfelder2009backbone}, 
the global network of ownership has a bow-tie structure with a very concentrated core of financial companies \cite{vitali2011network} -- a community structure that reflects geopolitical blocks \cite{vitali2014community}, while the embedding into geographical space explains several of the network properties \cite{vitali2011geography}. 
The global ownership network appears to be very resilient in terms of the power structure, even to the dramatic events of the 2008 financial crisis \cite{glattfelder2019global}.

One stream of work since the early days of the discipline has focused on the structure of the network of credit contract among financial institutions (hereafter interbank credit network). These networks are the natural and simplest empirical counterpart for many of the financial contagion models (see Section \ref{sec:dyn}).   
However, data on these type of relationships is confidential and available only to or through financial supervisors. 
In the very recent years, some supervisors have gained access to data that correspond to daily multiplex, cross boundary networks (e.g.\ data on derivative contracts collected in Trade Repositories\cite{derrico2018does,bardoscia2019multiplex}). However, in the beginning, only static data was available for individual countries, due to boundary issues. Empirical papers have analysed interbank networks in several countries. The early works on Austria \cite{boss2004network} and US \cite{soramaki2007topology}, paved the way to a stream of works covering e.g.\ 
Brazil \cite{Cajueiro2008,e2010brazilian}, Belgium \cite{degryse2007interbank}, Colombia\cite{leon2014rethinking}, Germany \cite{craig2014interbank}, Italy \cite{demasi2006fitness,iori2008overnight,iazzetta2009topology,finger2013network}, Japan \cite{imakubo2010transaction}, Mexico \cite{martinez2014empirical}, Switzerland \cite{muller2006interbank},US \cite{demiralp2006overnight,bech2010topology}.
Overall, the above works have highlighted in national banking systems the existence of some stylized facts, i.e.\ statistical features that are common across the different networks. These financial networks were found to be typically very sparse, with heavy-tailed degree distributions, high clustering and short average path length, being disassortative.

A specific strand of the literature has focused on characterising the underlying topology of those networks. A few studies looked for the most appropriate emergent block description of financial networks, whether they feature a subset of tightly-connected institutions, the \emph{core}, and a subset of institutions, the \emph{periphery}, which are loosely connected among each other and often connected to the core\cite{van2014finding,gabrieli2014network,fricke2015core,silva2016financial,kojaku2018structural}. 
Post-crisis, the establishment of central clearing counterparties (CCPs) has meant that many contracts were rerouted through a single institution (the CCP).
The potential benefits and risks of such change in the network topology have been investigated in \cite{cont2014central,duffie2011does,cont2016credit,duffie2015central,heath2016ccps,markose2017systemic,poce2018what}.

\subsection{Multiplex and higher-order networks}
The financial networks discussed so far condense all the information about the relationships between a pair of financial institutions into one (weighted) edge. 
This is often a useful abstraction, but in realty those relationships are more complex and this can matter for the propagation of risk. 
Multiplex networks\cite{bianconi2018multilayer,battiston2018multiplex} provide a natural framework to describe such relationships and 
first empirical studies have shown that different layers are not structurally similar. 
Examples include: credit and liquidity exposures in the UK interbank market\cite{langfield2014mapping}, payments and exposures in the Mexican banking system\cite{martinez2014empirical}, the Mexican interbank market\cite{molina2015multiplex}, the Italian interbank market\cite{bargigli2015multiplex}, Colombian financial institutions and market infrastructures\cite{berndsen2016financial}, the EU derivatives market\cite{abad2016shedding}, the UK interest rate, foreign exchange, and credit derivatives market\cite{bardoscia2019multiplex}, corporate networks\cite{jeude2019multilayer}.
Several works have focused on financial contagion on multiplex networks, see Section \ref{subsec:contagion_multiplex}.

Another fruitful area of investigation has been the network implied by the derivative market.
Due to data availability, most early studies have focused on 
credit default swaps (CDSs), derivative contracts in which an institution offers to insure another institution over the default of a third institution.
As such, they are an example of three-body interactions.
The existence of this market should allow institutions to hedge their risks and provide a solid market valuation of the financial risk of different market players. 
Nevertheless, CDS markets, as shown in a series of financial networks studies \cite{heise2012derivatives,brunnermeier2013assessing,roukny2014network,abad2016shedding}, can themselves become a channel of contagion in the possible situation where CDS insurers absorb too much risk upon themselves \cite{derrico2018does}.

\subsection{Correlation- and similarity-based networks}\label{subsec:correl}
\begin{figure}
\includegraphics[width=0.95\columnwidth]{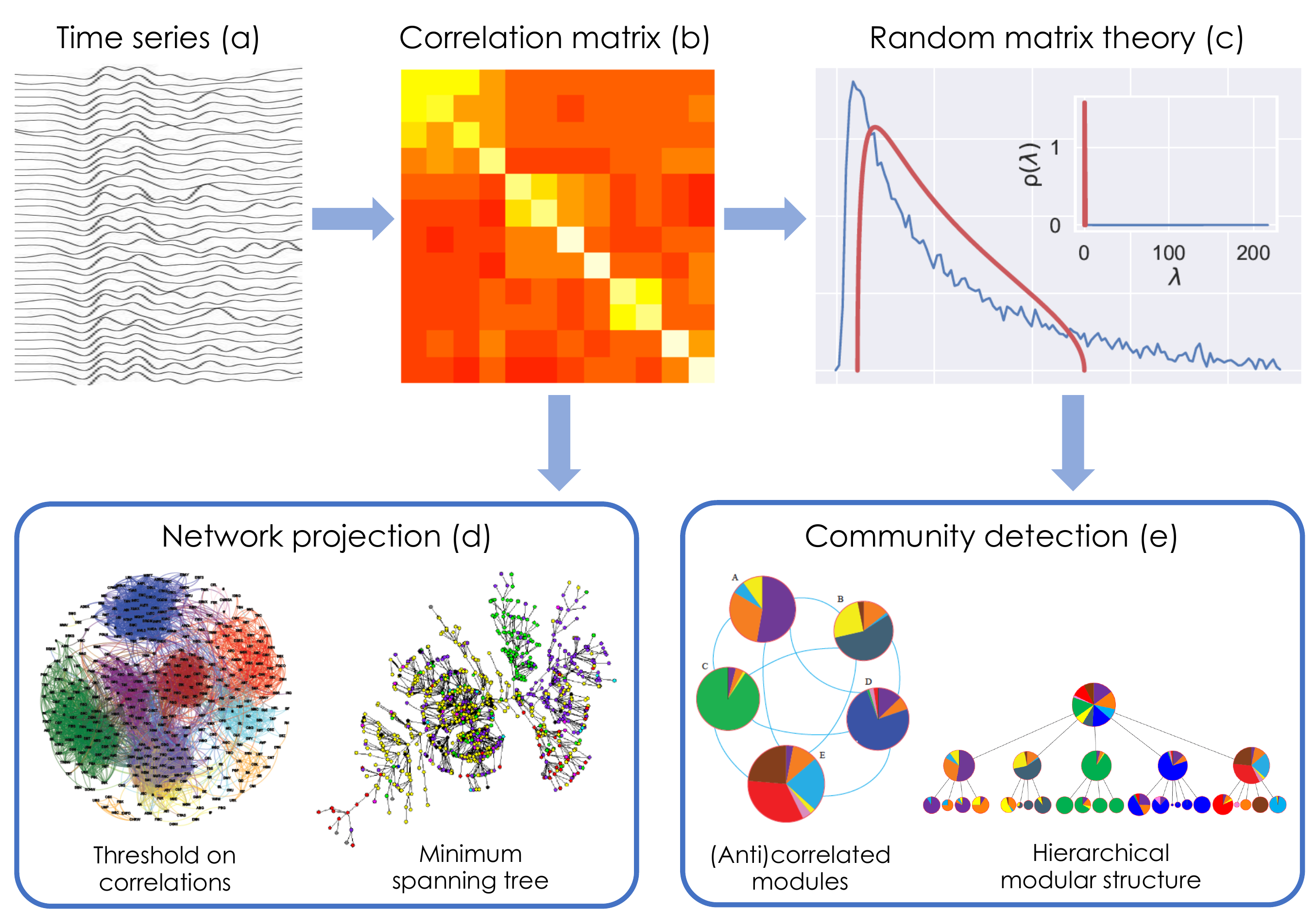}
\caption{Procedure illustration for the analysis of network structures and their communities out of time series data.}
\label{fig:diego}
\end{figure}

In several circumstances, financial entities are not necessarily related via `direct' interactions (such as flows of money, holdings of shares, or financial exposures), but via some possibly indirect form of correlation or similarity. 
Technically, correlation- and similarity-based networks are \emph{one-mode projections} of a set of multivariate time series (see Figure \ref{fig:diego}a-b) or of a \emph{bipartite} network, i.e.\ a network where connections exist only between nodes of two different types $A$ and $B$ (e.g.\ companies and directors, with a director being connected to a company if he/she sits in the board of that company). A one-mode projection contains only nodes of one type (say $A$) and any two such nodes are connected to each other with an intensity proportional to their pairwise temporal correlation or to the number of their common neighbours of the other type in the original bipartite network (for instance, two directors are connected by a link indicating the number of common boards in which they sit).
Various distinctive features of correlation-based networks require special caution and can make their analysis more complicated than that of other types of networks, as we explain below.
\begin{itemize}
\item While other types of networks can be arbitrarily sparse, matrices obtained from empirical correlation, similarity or (Granger) causality~\cite{billio2012econometric} generally do not contain zeroes, so they do not immediately result in a network (unless one interprets one such matrix as a trivial, fully connected graph with weighted links). This property has led to the introduction of several filtering techniques aimed at sparsifying those matrices while retaining the ``strongest’’ connections.
\item In presence of heterogeneous entities, the same measured correlation value might correspond to very different levels of statistical significance for distinct pairs of nodes. For this reason, simply imposing a common global threshold on all correlations is inadequate, and alternative filtering techniques that project the original correlation matrix onto minimum spanning trees~\cite{MantegnaCB}, maximally planar graphs~\cite{TumminelloCB}, or more general manifolds~\cite{DiMatteoCB} have been introduced (see Figure \ref{fig:diego}d). These approaches found that financial entities belonging to the same nominal category can have very different connectivity properties (e.g.\ centrality, number and strength of relevant connections) in the network~\cite{BonannoCB}. A question left open is the theoretical justification for the choice of the embedding geometry wherein the network is constructed.
\item Generally, all entries of empirical similarity matrices tend to be shifted towards large values, as a result of a common overall relatedness existing among all nodes, as for example a market trend.
This ``global mode’’ obfuscates the genuine dyadic dependencies that any network representation aims at portraying.
\item The measurement of correlation matrices is intrinsically prone to the \emph{curse of dimensionality}: in order to measure with statistical robustness the $n(n-1)/2$ entries of a correlation (or similarity) matrix, one needs a number $m\ge n$ of temporal observations (or features) in the original data to avoid dependency and statistical noise. 
Unfortunately, increasing $m$ for a given set of $n$ nodes is often not possible in practice, for instance because one would need to consider a time span so long that nonstationarities would unavoidably kick in, making the measured correlation unstable and not properly interpretable.
\end{itemize}

The above complications require the comparison with a proper null hypothesis that controls simultaneously for node heterogeneity, for a possible ``obfuscating’’ global mode and for ``cursed’’ noisy measurements. An important caveat here is that in correlation-based networks \emph{even the null model necessarily has mutually dependent links}. This  key difference arises from the fact that, if node $i$ is positively correlated with (or similar to) node $j$, which is in turn positively correlated with node $k$, then nodes $i$ and $k$ are typically also positively correlated. This ``metric’’ constraint survives also in the null hypothesis of random correlations, while it does not apply in usual null models developed for networks. Therefore, naively using those null models introduces severe biases in the statistical analysis of correlation-based networks. Adequate null models can be defined in terms of a random correlation matrix~\cite{BouchaudCB,StanleyCB,LilloCB,MacMahonCB,BaruccaCB} (technically, a Wishart matrix~\cite{RandomMatricesCB}), whose entries are automatically dependent on each other in the desired way. Indeed, random matrix theory~\cite{RandomMatricesCB,RandomMatrices2CB} has become a key tool in the analysis of correlation matrices. A successful use of this theory is the comparison of the spectra of empirical and random correlation matrices and the selection of the empirically deviating eigenvalues to construct the filtered (non-random) component of the measured matrix~\cite{MacMahonCB} (see Figure \ref{fig:diego}c). This filtered matrix enables the detection of patterns such as communities (see Figure \ref{fig:diego}e) and the identification, in a purely data-driven fashion, of empirical dependencies that, again, turn out to be unpredictable \emph{a priori} from the nominal classification or taxonomy of nodes.

Research in this direction is currently very active
\cite{AnagnostouCB,BaruccaCB,ahelegbey2016bayesian,ahelegbey2016sparse}: data-driven groups of strongly correlated stocks~\cite{MacMahonCB,AlmogCB} and CDSs~\cite{AnagnostouCB} have been found to be unpredictable from industry or sector classification and to improve the performance of standard factor models for risk modelling and portfolio management~\cite{AnagnostouCB}. Theoretically, more general matrix ensembles have been studied using notions from supersymmetry~\cite{BaruccaCB} to further refine the analytical characterization of the null hypothesis.\\

\begin{figure}
\centering
\includegraphics[width=0.95\columnwidth]{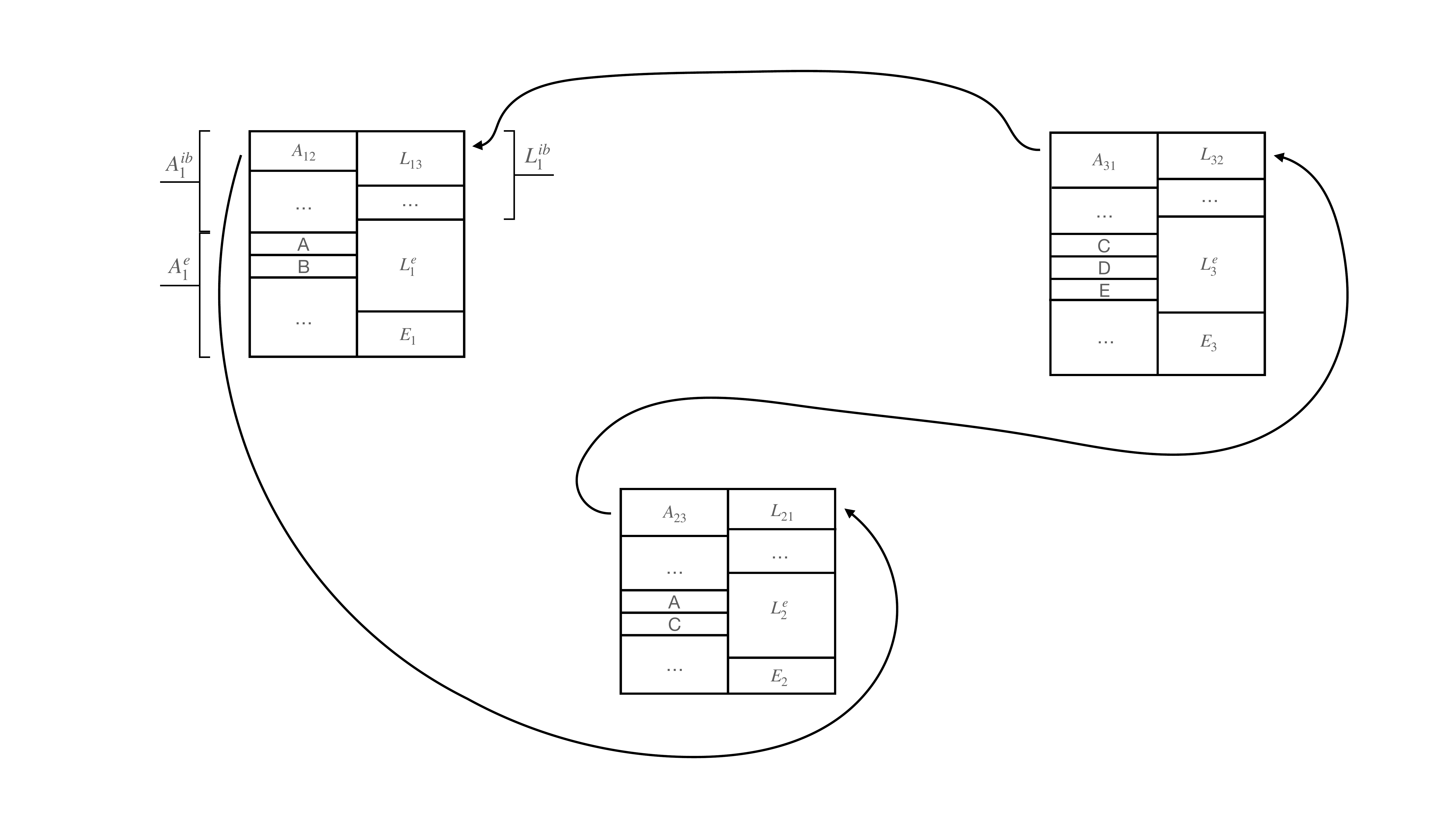}
\includegraphics[width=0.95\columnwidth]{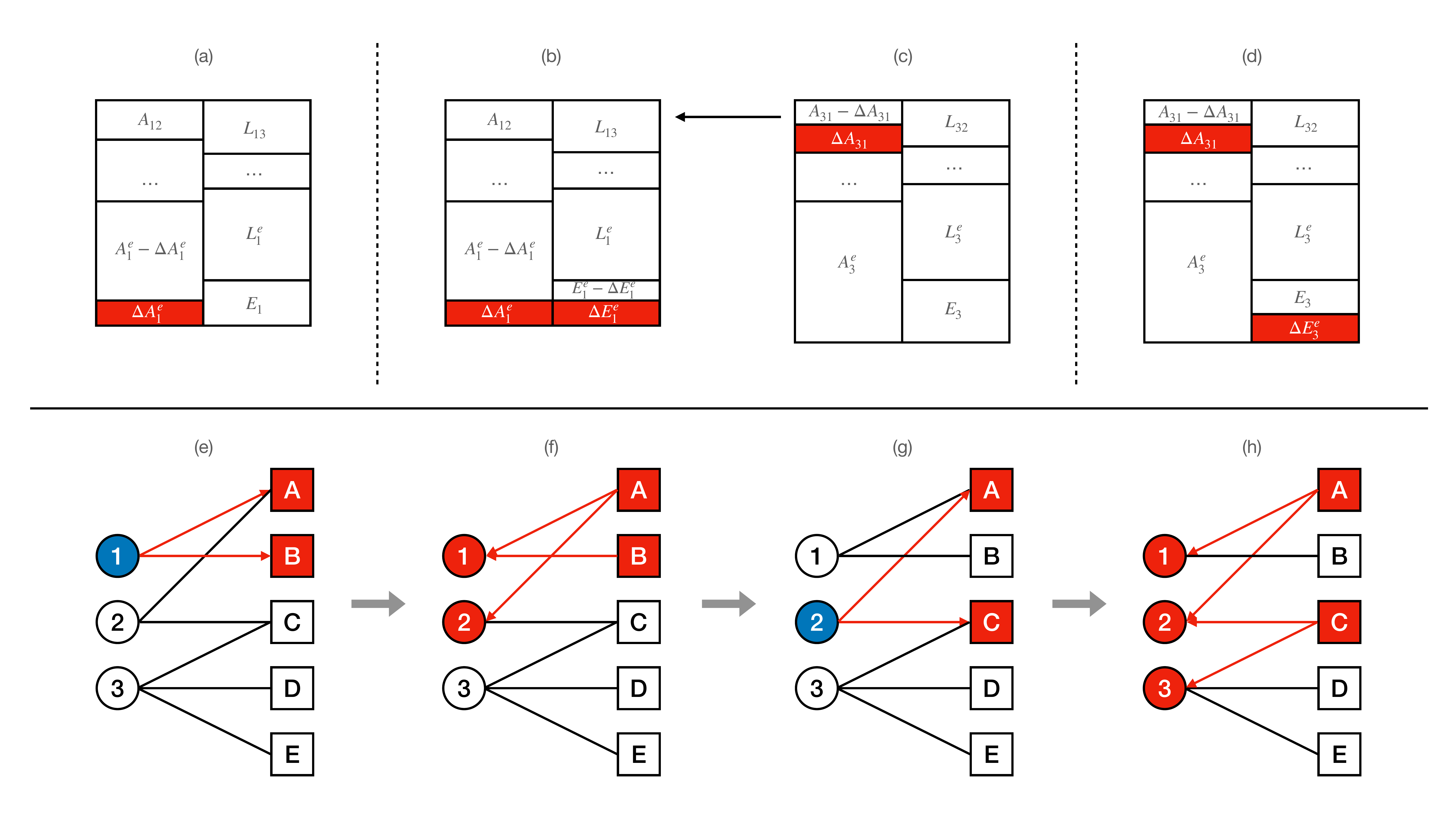}
\caption{\textbf{Interbank networks and their dynamics.} In the top panel we show a stylised interbank network composed by three banks, each represented by its balance sheet. On the asset side we have interbank $A_i^{ib}$, further broken down in individual exposures (e.g.\ $A_{12}$ is the exposure of bank 1 to bank 2), and external assets (e.g.\ A, B, $\ldots$). On the liability side we have interbank liabilities $L_i^{ib}$, similarly broken down, external liabilities $L_i^e$, and equity $E_i$. The bottom panels show two possible contagion channels, solvency contagion and overlapping portfolios. In the former, an exogenous shock hits external assets of bank 1 (a) and is absorbed the bank 1's equity (b). Since bank 2 is exposed to bank 1, it re-valuates its interbank asset $A_{31}$ (c). The exact valuation method depends on the specific model. Finally, the reduction in bank 2 assets is absorbed by its equity (d). In the latter, bank 1 sells assets A and B, for example to meet its leverage target (e). This causes A and B to depreciate. Asset values of banks 1 and 2, which hold A and B, are reduced (f). As a consequence, bank 2 now needs to deleverage and sells assets A and C (g). Those assets depreciate and asset values of banks 1, 2, and 3 are reduced (h).}
\label{fig:interbank}
\end{figure}

\section{Dynamics of Financial Networks} \label{sec:dyn}
\subsection{Direct contagion: solvency and liquidity} \label{subsec:dyn_bilat}
In this section we will review models of financial contagion that focus on bilateral relationships between financial institutions (for brevity referred to as \emph{banks} in the following),
which are one of the most common example of financial networks.
Most models can be grouped under the general framework illustrated in the top panel of Figure \ref{fig:interbank}. The idea is that to each bank are associated some dynamical state variables representing key quantities in their balance sheet. 
Those variables are updated via dynamic equations that depend strongly on the relationships between banks, which are typically static.

As said before, each bank is represented by its balance sheet, which consists of an asset side (things that generate income for the bank, such as loans extended to households, to other banks or firms), and a liability side (claims of other economic agent towards that bank), such as customer deposits, funds borrowed from other banks or firms, bonds and shares issued.
The balance sheet identity prescribes that the sum of assets of each bank is equal to the sum of its liabilities. Liabilities have different priorities (the ``seniorities'' introduced above). In case a bank fails, its assets are liquidated and its liabilities are paid back starting from those with a higher priority. The liability with the lowest priority is the equity, which corresponds to the residual claim of shareholders after all other liabilities have been paid back. Therefore, it is a measure of the bank's net worth. 
Both assets and liabilities can be split according to the market they belong to, in particular it is customary to distinguish between \emph{interbank} (or \emph{network}) and \emph{external}. The interbank liabilities of bank $i$ are the obligations that $i$ has to other banks in the system, such as payments to be made imminently, or loans which correspond to payments to be made in the future. Similarly, the interbank assets of bank $i$ are the obligations that other banks in the system have towards $i$.
To each interbank liability (e.g.\ of $i$ towards $j$) it corresponds an interbank asset (of $j$ towards $i$). Hence, interbank assets and liabilities are equivalent representations of the obligations between pairs of banks.
The network is built simply by associating to each bank one node and to each interbank liability (or asset) one link.
Those networks are \emph{directed} (obligations are not necessarily symmetric), \emph{weighted} (by the monetary amount of the obligation), and \emph{without loops} (banks do not make payments to themselves). 
For the sake of brevity, here we have considered the case in which all obligations between each pair of banks are aggregated into one single interbank liability (or asset). More granular models based on multiplex networks can overcome this limitation, see Section \ref{subsec:contagion_multiplex}.

The usual approach followed is to hit one or more banks with an exogenous shock, for example by reducing the value of their external assets, and to propagate such shock across the network, not unlike an epidemic. 
Formally, this corresponds to a dynamical process (triggered by an external shock) on the network that allows balance sheet variables to evolve.
This approach is conceptually similar to the stress tests
of the banking sector as implemented worldwide by regulators after the 2008 financial crisis. 
However, those usually consider banks in isolation and neglect interactions between them. 
Therefore, the natural policy application of these models has been to incorporate network effects into more traditional stress testing models.

Although specific models differ in their implementation details, they describe only a handful of basic shock propagation mechanisms. 
Firstly, we may have {\em liquidity contagion}.
In this case, the relevant balance sheet quantities are interbank liabilities, which represent payments to be delivered imminently, and liquid assets, which are a subset of external assets and consist of cash or assets that can readily be converted into cash. 
Some banks are able to pay their obligations in full, if the sum of their liquid assets and payments received by other banks exceed their payment obligations. 
Banks that are not able to do so will make reduced proportional payments, as in \cite{eisenberg2001systemic} or in \cite{rogers2013failure}, which accounts also for bankruptcy costs, or no payment at all, as in \cite{bardoscia2019full}. 
Contagion spreads when there are banks that would have been able to meet their own obligations if they had received their incoming payments.
However, since some of those payments were not (or were only partially) delivered, they are not able to fully deliver their own payments, potentially putting banks on the receiving end of their payments in the same situation. 
More recently, a few studies have focused on liquidity shocks originating from the derivatives market, see e.g.\ \cite{cont2016credit,paddrik2020contagion,bardoscia2019simulating}.

Secondly, we can have {\em solvency contagion}. 
Solvency contagion occurs when the insolvency or the reduction in creditworthiness of a bank has an effect on its creditors.
The simplest form of solvency contagion is known as \emph{contagion on default}.
This means that when bank $i$'s equity becomes smaller than or equal to zero, $i$'s creditors  will write-off their interbank assets towards $i$ because they do not expect to be fully paid back. 
(Negative equity is a common sufficient condition for insolvency or default. However, resolution frameworks put in place after the 2008 financial crisis imply that banks can be wound down when they fail to comply with regulatory requirements, even though their equity is positive.)
In the most conservative case $i$'s creditors set the value of the corresponding interbank assets to zero, as they expect to recover nothing from the defaulted bank \cite{cont2013network}. 
In general, they will discount their interbank assets by a coefficient between zero and one known as recovery rate, such as in \cite{furfine2003quantifying}.
When $j$, one of $i$'s creditors, writes-off its interbank assets, the total value of $j$'s assets is reduced and, via the balance sheet identity, also the total value of $j$'s  liabilities. 
Having the lowest priority, in the first instance it is $j$'s equity to absorb the losses.
However, if $j$'s equity is not large enough, it will default too, thereby triggering write-offs by its own creditors.
The spreading of defaults across the financial network is known as \emph{default cascade} or as \emph{domino effect}.
These models are mathematically similar to linear threshold models and are therefore amenable to analytical treatment \cite{may2010systemic}, for example to derive the size of the default cascade \cite{gleeson2012systemic,amini2016resilience}.
In more general models, write-offs are triggered not only by defaults, but also by increases in probabilities of default. 
This is the approach followed by the family of DebtRank models \cite{battiston2012debtrank,bardoscia2015debtrank,bardoscia2016distress}, by empirical models \cite{fink2016credit}, or by valuation models \cite{elsinger2006risk,fischer2014no-arbitrage,barucca2020network,bardoscia2019forward} (see also the bottom panel of Figure \ref{fig:interbank}).
Such mechanism mimics the accounting requirement of marking asset to market, which has been a large source of losses during the 2008 financial crisis, see e.g.\ \cite{basel2011capital}.

Thirdly, we have {\em funding contagion}. 
This occurs when banks that have previously lent to bank $i$ decide not to renew their loans once they expire \cite{gai2011complexity,brandi2018epidemics}. 
Similarly to solvency contagion, the decision can be triggered by a change in the creditworthiness of bank $i$.
See \cite{cimini2016entangling} for a model that integrates solvency and funding contagion.

Models of bilateral exposures have been also used to investigate the relationship between the underlying topology of the network and its stability. 
The early works \cite{allen2000financial} and \cite{freixas2000systemic}, following standard economic theory, show that more diversified (and therefore more interconnected) networks are more resilient, as shocks are dispersed across more banks. 
However, \cite{battiston2012credit} argue that this is not always true and 
in \cite{acemoglu2015asystemic} it is shown that a more interconnected network is more resilient only for small shocks, but that it is actually less resilient for large shocks, in line with the intuition that financial networks might be ``robust-yet-fragile'' \cite{haldane2009rethinking}. 
Similarly, in \cite{gai2011complexity} it is shown that, while the probability of widespread contagion can be small, systemic events can be very severe. 
Both \cite{nier2007network} and \cite{elliott2014financial} show that diversification does not have a monotonous effect on the extent of default cascades.
Also, the role of the topology is crucial\cite{kobayashi2013network}, with no single network architecture superior to the other ones\cite{roukny2013default}. 
In \cite{bardoscia2017pathways} the instability of the contagion dynamics (see also \cite{markose2012too}) is linked to the presence of specific topological structures (unstable cycles), which are likely to appear in a more diversified network. 
A different approach is followed by \cite{glasserman2015likely}, in which the relationship between interconnectedness and resilience is investigated with minimal information about the network structure.

Yet another application consists in assessing and designing (optimal) policies. 
For example, \cite{batiz2016calibrating} show that limits on exposures often, but not always, reduce systemic risk and \cite{huser2018systemic} develop a toolkit to test the impact of bail-ins. 
Several studies focus on the impact on public finances: in \cite{capponi2016capital} it is shown that resolution frameworks can be effective in reducing bail-out, \cite{Alter2015} investigates centrality-based bail-outs, while \cite{minca2014optimal} and \cite{capponi2020optimal} determine conditions for optimal bail-outs. 
Closely related are the studies on the controllability of financial networks \cite{delpini2013evolution,galbiati2013power}, aimed at reducing systemic risk, e.g.\ by a targeted tax \cite{poledna2016elimination}, by requiring banks to disclose their systemic impact \cite{thurner2013debtrank}, or by explicitly optimising the exposures\cite{diem2020minimal}.

\subsection{Indirect contagion: overlapping portfolios} \label{subsec:dyn_overlap}
Shocks can propagate between banks even if they are not directly connected through a contract. This happens for instance when they invest in common assets. 
If one bank is in trouble, it may choose to sell some of its assets.
This would causes their devaluation and therefore losses for the other banks that had invested in the same assets, which may cause these banks to sell their assets in turn, and so on. Although this type of contagion is mediated by the market through the price, interactions can still be modeled as a network of overlapping portfolios. 
The network is a bipartite one, with two types of nodes representing banks and assets, and links connecting banks to the assets in their portfolio (see also bottom panel of Figure \ref{fig:interbank}), so it can be considered a particular case of the correlation networks described in Section \ref{subsec:correl}, despite the fact that in this case we are interested simultaneously in both partitions since the overlap change dynamically. 

Similarly to the case of direct exposures, the goal is that of understanding how the properties of the overlapping portfolio network affects its stability, and under what conditions the system is able to absorb vs.\ amplify exogenous shocks \cite{huang2013cascading, caccioli2014stability,greenwood2015vulnerable,corsi2016micro,duarte2018fire,cont2017fire}.
In order to model the dynamics of shock propagation on this network, one needs to specify how banks react to losses in their portfolios (e.g.\ how they readjust their portfolios to manage risk) and how asset prices react to banks' trading activity.
The response of prices to liquidation is typically implemented by means of a market impact function \cite{bouchaud2009markets} that links the liquidation volume of an asset to its price: The more an asset is being liquidated, the higher its devaluation. 
Most of the literature considers market impact functions that are linear in returns or log-returns, although more complex forms that take into account the fact that when an asset is largely devalued other investors would step in the attempt to buy the asset at a cheap price have also been considered \cite{cont2017fire}. 

Concerning the dynamics of banks, the simplest choice is that of a linear threshold model \cite{watts2002simple}, where a bank is passive as long as its losses remain below a given threshold (typically chosen to be equal to its equity), and liquidate its entire portfolio otherwise \cite{huang2013cascading,caccioli2014stability}. Under this assumption, it is possible to derive analytical results for the case of random networks, when the dynamics can be approximated by a multi-type branching process \cite{caccioli2014stability}. 
When the branching process is supercritical, even a small exogenous shock can propagate throughout the network. 
Using this analogy, it is possible to identify regions in the parameter space (average degree of banks and assets in the network, strength of market impact, leverage) where cascades of defaults occur, and to show that increasing diversification -- which reduces the risk of individual institutions -- does not necessarily increase systemic stability \cite{caccioli2014stability}.
The study of random networks is important from the theoretical point of view, but the ultimate goal of contagion models is that of characterizing the stability of realistic systems. 
For instance, \cite{huang2013cascading} considers a network of overlapping portfolios between US commercial banks.
A stress test is carried out by means of numerical simulations, showing the existence of phase transition-like phenomena between stable and unstable regimes.

While the assumption of a passive investor is a useful benchmark against which to assess the effect of active risk management --  and it may be realistic during a fast-developing crisis when banks do not have time to react before they default -- in practice banks would react to changing market conditions by actively rebalancing their portfolios. This happens because of internal risk management procedures, or because of regulatory constraints that need to be satisfied \cite{adrian2010liquidity}.
Active risk management is typically modeled by means of leverage targeting dynamics \cite{caccioli2014stability,greenwood2015vulnerable,duarte2018fire,corsi2016micro}, according to which banks that experience losses liquidate a fraction of their investment in the attempt to keep their leverage constant. In fact, it can be shown that leverage targeting is the optimal strategy of an investor who tries to maximize its expected return on equity while being subject to a VaR or Expected Shortfall constraint \cite{shin2010risk}. 
A dynamic in-between thresholding and leverage targeting is considered in \cite{cont2017fire}: Banks do not react until their losses exceed a given threshold, after which they target leverage. 
A stress testing exercise is performed on a network of overlapping portfolios between European banks, and the amount of indirect exposures induced by the network is quantified. 
Indirect exposure here refers to the fact that, through the network of overlapping portfolios, banks may be effectively (and unknowingly) exposed to assets they are not investing in.


Contagion due to overlapping portfolios also affects financial institutions other than banks. Recent studies have focused on funds, which also actively manage their portfolio by liquidating their portfolios in time of distress.
Refs.\ \cite{guo2016topology} and \cite{braverman2018networks} provide for instance empirical characterizations of the network of overlapping portfolios between funds, while \cite{fricke2017vulnerable} and \cite{baranova2017simulating} introduce stress testing frameworks to study the stability of US mutual funds and European investment funds. A study of US mutual funds is carried out in\cite{delpini2019systemic}, showing that the vulnerability of the network is larger compared to the benchmark of random networks where nodes' degrees are preserved. 

\subsection{Contagion on multiplex networks} \label{Contagion on multiplex networks} \label{subsec:contagion_multiplex}

As mentioned in Section \ref{subsec:dyn_bilat} granular models in which exposures are disaggregated, by maturity\cite{kusnetsov2019interbank}, seniority\cite{fischer2014no-arbitrage} (i.e.\ the order of repayment in the event of a sale or bankruptcy), or asset class\cite{feinstein2019obligations} can be represented by multiplex networks.
More specifically, in\cite{brummitt2015cascades} it is shown that mixing debts of different seniority levels makes the system more stable.
In practice, risks associated with different layers could either offset or reinforce.
In\cite{poledna2015multilayer} a multiplex representation of the Mexican banking system between 2007 and 2013 and is built an analysed.
Crucially, it is shown that focusing on a single layer can underestimate the total systemic risk by up to 90\% and that risks generated by individual layers cannot be simply summed, but interact among themselves in a non-linear fashion. 
A similar result is found by\cite{montagna2016multilayered} by using an agent-based model of the multiplex interbank network for large EU banks.

In Sections \ref{subsec:dyn_bilat} and \ref{subsec:dyn_overlap} we have seen that there are different contagion channels through which stress can propagate between financial institutions. 
Since each channel can be represented as a network, a complete characterization of financial contagion should therefore consider multiple contagion channels at the same time on a multiplex network. 
For example, in\cite{bookstaber2016looking} the possible interplay among layers corresponding to short‐term funding, assets, and collateral flows are outlined, clarifying how risk propagates from one layer to another using the case of Bear Stearns during the financial crisis.

The stability properties of the system depend on the interplay of the contagion processes across layers, which can differ in nature depending on the type of layer. 
By looking at a single layer at a time, one can fail to detect instability and fail to identify the possible contagion channels.
In\cite{burkholz2016systemic} it is found that, when two layers are not coupled weakly, systemic risk is larger in a multiplex network than in the aggregated single-layer network.
Moreover, the sharp phase transition in the size of the cascade is more pronounced in the multiplex network.

Some early works on financial contagion due to counterparty risk \cite{gai2010contagion,nier2007network,cifuentes2005liquidity,may2010systemic} consider in fact the effects of fire sales by assuming the existence of one asset that is common to all banks and is liquidated when banks default, but their focus remains on understanding how the topology of the interbank exposures network affects its stability.
More recently, by using data on direct interbank exposures between Austrian banks and also by assuming the existence of a common asset between banks, a study in\cite{caccioli2015overlapping} suggests that the interaction between contagion channels can significantly contribute to aggregate losses. 
These findings are confirmed by \cite{poledna2018quantification} using detailed data on direct exposures and overlapping portfolios between Mexican banks.

\section{Statistical physics of financial networks}

\begin{figure}
\centering
\includegraphics[width=0.75\linewidth]{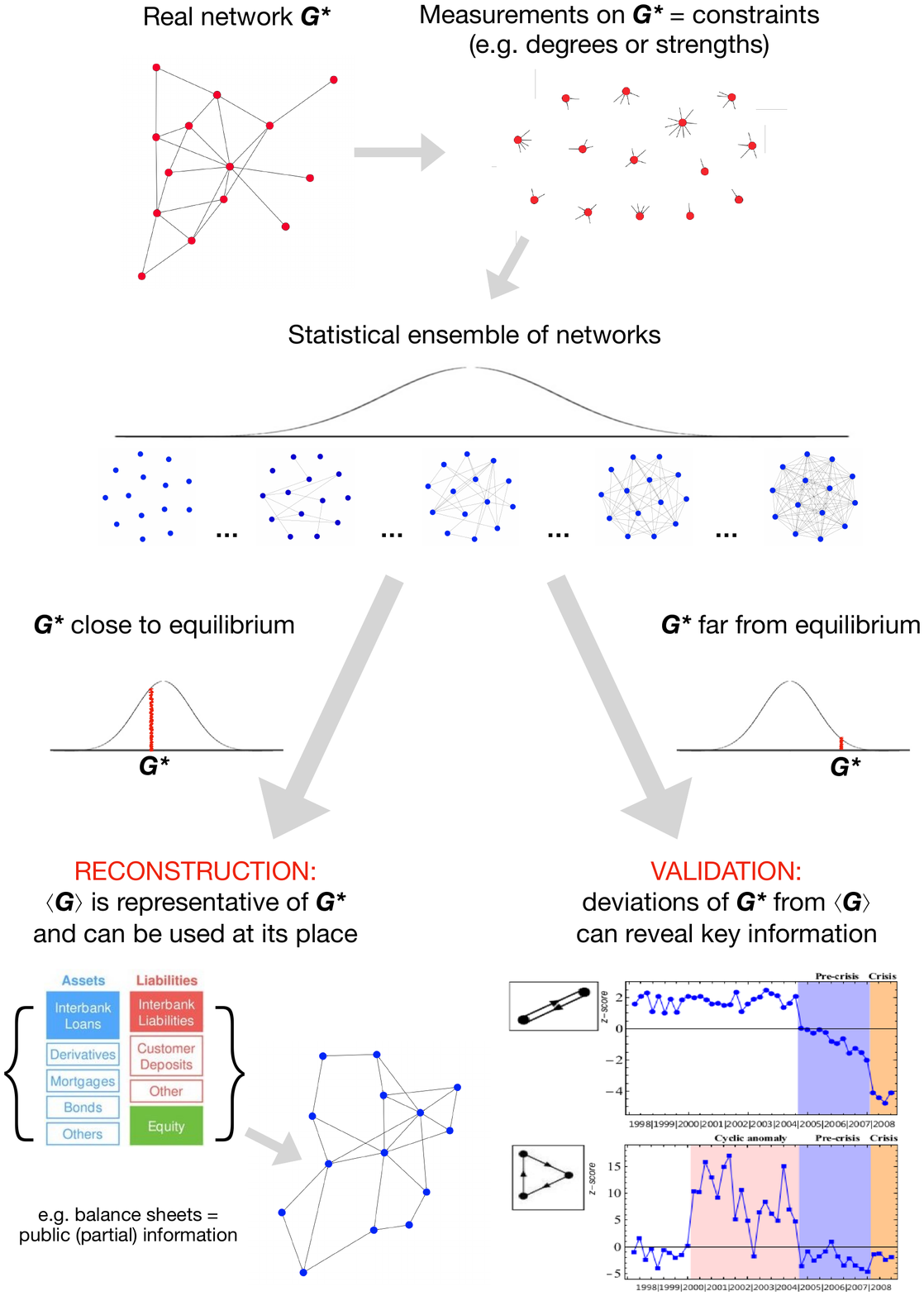}
\caption{Statistical physics approach to modeling networks at equilibrium and out-of-equilibrium. Both cases employ the formalism of the canonical ensemble in statistical physics using a double optimization step: first, the constrained maximization of Shannon entropy determines the functional form of the probability distribution $P(\mathbf{G}^*|\mathbf{\theta})$; then, the maximization of the likelihood functional $\mathcal{L}(\mathbf{\theta})=\ln P(\mathbf{G}^*|\mathbf{\theta})$ provides the recipe to numerically estimate the parameters $\mathbf{\theta}$, ensuring that the expected degree sequence matches the empirical one. When the desired features of a network are reproduced by the model one speaks of \emph{network reconstruction}; otherwise, the model can be used to test their statistical significance.}
\label{fig7}
\end{figure}

\subsection{Networks at equilibrium} \label{sec:equilibrium}
The dynamical models discussed in the previous section assume that the \emph{presence} and \emph{magnitude} of relationships between financial institutions is known.
Unfortunately, due to confidentiality issues, that information is often only accessible to regulators. Even then, regulators only have a partial view of the financial network, typically limited to their jurisdiction.
The present section reviews the attempts that have been made so far to overcome this difficulty. More importantly, knowledge of a snapshot of the financial system, partial or complete could complement the various models in predicting the system evolution. 

The problem of partial information is of course very general in physics; even the more particular  problem of \emph{missing} or \emph{partial information} about network structures affects a wide range of real-world systems (e.g.\ metabolic or ecological webs whose interlinkages patterns are only partially accessible due to experimental limitations) and has led to the birth of a research field known as \emph{network reconstruction} \cite{squartini2018reconstruction,cimini2019statistical}.
For this reason, many reconstruction algorithms have been proposed so far (see 
Table \ref{tab_meth} for a description of the most representative attempts). In all these methods we either deal with \emph{deterministic} or rather with the \emph{probabilistic} character of the reconstruction procedure. Among the ones belonging to the first class, the \emph{MaxEnt} \cite{wells2004financial}, the \emph{Iterative Proportional Fitting} \cite{upper2004estimating,mistrulli2011assessing} and the \emph{Minimum Density} \cite{anand2015filling} algorithms deserve to be mentioned. An evident limitation affecting these deterministic methods is that they produce a single instance of a network, thereby giving zero probability to any other configuration (often including the true unobserved network) \cite{parisi2020faster}.  The same conclusion holds true also for methods combining a probabilistic recipe for the topological estimation (i.e.\ for determining the links presence) with a deterministic recipe for the weights estimation \cite{gandy2016bayesian}.
To overcome the previous limitations fully probabilistic (statistical) methods have recently been introduced. The most effective approaches to the network reconstruction problem are the ones rooted in statistical physics and rest upon the so-called \emph{maximum entropy principle} \cite{shannon1948mathematical,jaynes1957information,park2004statistical,squartini2011analytical} (see Box 2). Loosely speaking, the recipe is to \emph{maximize the uncertainty} about the system at hand once the available information is properly accounted for. This approach prevents making assumptions that are not supported by empirical information and that would otherwise bias the entire estimation procedure.
While uncertainty maximization is carried out by maximizing the \emph{Shannon entropy}, the available information is included as \emph{constraints} in the optimization procedure via the formalism of Lagrange multipliers. Therefore, after having identified the accessible network properties (usually represented by aggregate information, that in our case could be for example the total interbank lending and borrowing of each bank) these methods assume that the structure of the entire network can be simply explained in terms of the selected properties. 

The first entropy-based algorithms have been based on the assumption that the constraints concerning the binary and the weighted network structure \emph{jointly} determine the reconstruction output (e.g.\ as in the \emph{Enhanced Configuration Model}, simultaneously constraining the degrees and the strengths of nodes \cite{mastrandrea2014enhanced,gabrielli2019grand}). However the inaccessibility of empirical degrees (i.e.\ number of lenders or borrowers of each bank) makes these methods inapplicable for reconstructing financial networks. This has led to the introduction of two-step algorithms \cite{cimini2015estimating,cimini2015systemic} that perform a preliminary estimation of the node degrees using the \emph{fitness model} \cite{caldarelli2002scale} to overcome the lack of binary information. In particular the \emph{density-corrected Gravity method} (dc-GM) \cite{cimini2015systemic} has been tested in four independent `horse races' \cite{anand2017missing,mazzarisi2017methods,ramadiah2017reconstructing,lebacher2019search} and found to systematically outperform competing reconstruction methods.

The idea of estimating the weighted structure of a given network \emph{conditional} on the preliminary, binary estimation step has been properly formalised only recently \cite{parisi2020faster}. The score function to optimize in this case is the \emph{conditional Shannon entropy}, whose maximization is constrained to reproduce the (purely) weighted information available on the system, by considering as \emph{prior} information the topological structure of the network -- be it empirical of inferred by a binary reconstruction method.

Importantly, besides confirming the success of a given reconstruction algorithm, the agreement between the \emph{empirical} trends and the \emph{expected} ones has deeper implications. Firstly, that the network to be reconstructed is close to the average (equilibrium) configuration of the canonical ensemble defined by the aforementioned constraints (see Figure \ref{fig7} for a sketched representation of this point). Secondly, that the network dynamics, driven by the dynamics of the constraints themselves, is \emph{quasi-stationary} \cite{squartini2015stationary}. These two consequences result in the fact that the real-world system under analysis is characterized by smooth structural changes rather than abrupt transitions. While smooth changes characterizing quasi-equilibrium networks can be generally controlled for, this is not possible in case of abrupt transitions, characterizing non-stationary networks.
Notice, however, that while being `at equilibrium' represents a \emph{necessary} condition to achieve an accurate reconstruction, it does not represent a \emph{sufficient} condition. In fact, a real-world network may still deviate from the (equilibrium) expectations provided by the chosen constraints; however, if the deviating patterns are moderate and similar in time (i.e.\ they appear as \emph{fluctuations} around the expected values) then the network can still be considered consistent with a quasi-stationary one, yet not completely driven by the dynamics of the chosen constraints. Hence, successfully reconstructing a network ultimately means that firstly one has to check the (out-of-)equilibrium character of its dynamics and, in case, secondly detect the right amount of information to provide an accurate description of it.

As illustrative examples, we consider two systems (an economic and a financial network). One of the most studied systems is the International Trade Network (ITN) whose nodes represent world countries and links represent export relationships between them \cite{serrano2003topology,fagiolo2009world,barigozzi2010multinetwork,fronczak2012structural,fronczak2012statistical,duenas2013modeling}. 
The evidence that many of its properties change considerably with time (e.g.\ the total number of nodes double across the period of fifty years ranging from 1950 to 2000 \cite{fagiolo2012null}) makes it an ideal system to test its (out-of-)equilibrium character.
As several contributions have revealed, the ITN structure is accurately reproduced by the local reciprocity, once ITN is represented as a binary, monopartite network (that can be undirected or directed, either in a monoplex or a multiplex fashion). This is particularly evident upon considering the so-called \emph{motifs} whose significance profiles remain quite stable in time \cite{squartini2015stationary}. Thus, even if the constraints may vary considerably across the history of the world trade -- presumably because of `hexogenous' effects as the creation of new states -- the deviations from the expectations induced by the former ones are bounded and systematic, hence making the quasi-equilibrium character of the ITN manifest. When coming to analyse the ITN weighted structure, the limitations of purely weighted constraints in successfully reproducing the corresponding quantities are overcome by employing the dc-GM or one of the conditional reconstruction algorithms recently proposed \cite{cimini2015systemic,parisi2020faster}.
The second example is provided by the Dutch Interbank Network (DIN) \cite{squartini2013early}. Differently from the ITN case, checking its consistency with maximum-entropy ensembles of graphs reveals it to be clearly out-of-equilibrium. It is, therefore, natural to question the usefulness of the aforementioned formalism in case non-stationary networks are met: the next section is devoted to answer this question.

\begin{table}\footnotesize
\caption{Overview of the various reconstruction methods that can be found in the literature. The column ``ME'' indicates whether the method is based on maximum entropy, ``Type'' denotes for the density of the reconstructed network and ``Category'' is either deterministic or probabilistic depending on whether the method generates a single network instance or an ensemble. See \cite{squartini2018reconstruction} for a detailed theoretical comparison.}
\begin{tabularx}{\textwidth}{lcllXc}
\toprule
{\bf Name} & {\bf ME} & {\bf Type} & {\bf Category} & {\bf Brief description} & {\bf Ref.} \\
\midrule\addlinespace
MaxEnt & \cmark & Dense & Deterministic & Maximizes Shannon entropy on network 
entries by constraining marginals & 
\cite{wells2004financial,upper2011simulation} \\
\addlinespace
IPF & \cmark & Tunable & Deterministic & Minimizes the KL divergence from MaxEnt 
& \cite{bacharach1965estimating} \\
\addlinespace
MECAPM & \cmark & Dense & Probabilistic & Constrains matrix entries to match, on 
average, MaxEnt values & \cite{digiangi2018assessing} \\
\addlinespace
Drehmann \& Tarashev & \cmark	& Tunable	& Probabilistic & Randomly 
perturbs the MaxEnt reconstruction & 
\cite{drehmann2013measuring} \\
\addlinespace
Mastromatteo et al. & \cmark & Tunable	& Probabilistic & Explores the space of 
network structures with the message-passing algorithm & 
\cite{mastromatteo2012reconstruction} \\
\addlinespace
 Moussa \& Cont & \cmark & Tunable & Probabilistic & Implements IPF on 
non-trivial topologies & \cite{moussa2011contagion} \\
\addlinespace
 Fitness-induced ERG 	 & \cmark	& Exact & Probabilistic & Uses the 
fitness ansatz to inform an exponential random graph model	 & 
\cite{cimini2015systemic,cimini2015estimating} \\
\addlinespace
Copula approach & \xmark & Dense & Deterministic & Generates a network via a 
copula function of the marginals & \cite{baral2012estimating} 
\\
\addlinespace
Gandy \& Veraart & \xmark	& Tunable & Probabilistic & Implements an 
adjustable Bayesian reconstruction & \cite{gandy2016bayesian} 
\\
\addlinespace
Montagna \& Lux & \xmark	& Tunable & Probabilistic & Assumes {\em ad-hoc} 
connection probabilities depending on marginals & 
\cite{montagna2017contagion} \\
\addlinespace
Ha\l{}aj \& Kok & \xmark & & Probabilistic & Uses external information to define a 
(geographical) probability map & \cite{halaj2013assessing} 
\\
\addlinespace
Minimum-Density & \xmark & Sparse & Probabilistic & Minimizes the network 
density while satisfying the marginals & \cite{anand2015filling} \\
\addlinespace\bottomrule
\end{tabularx}
\label{tab_meth}
\end{table}

\subsection{Networks out-of-equilibrium}\label{sec:Nooe}
In typical applications of statistical mechanics, a maximum entropy (null) model is used to obtain a probabilistic representation of a system, given the limited information available on it. This is precisely the spirit of the reconstruction methods mentioned in the previous section. Implicitly, the assumption is that the system is at the equilibrium. 

However, when we have full information on the network, we can use the entropy-based framework in the opposite way, i.e.\ to measure how much the system is out of the equilibrium (see Figure \ref{fig7}). The approach is to consider the null model as a particularly tailored benchmark to compare some non trivial quantities with: in this sense, everything that does not fit the null model highlights an aspect of the system we are studying that is not contained in the model itself. There is, of course, an important caveat: the information that we want to discount, i.e.\ the constraints in the maximisation of the entropy, should represent a very relevant information for the description of the system. Indeed, the sense of the out-of-equilibrium analysis is to investigate the properties of the network that cannot be explained simply by the constraints and it makes sense only if the constraints have some meaning for the representation of the phenomena under study\cite{billio2016entropy}.

An example may be useful to clarify the discussion. Let us consider the Dutch interbank network as analysed in reference~\cite{squartini2013early}. In this case, the full network was known, and the entropy-based null model were used to highlight the properties that were out of the equilibrium, or, otherwise stated, all the elements that were not compatible with the implemented null model. 
Before considering the comparison with a null model, the first observation was that the fraction of reciprocated links, i.e.\ the number of bank pairs that lend money to each other, experienced a drastic drop at the onset of the world financial crises. 
Such a measure is commonly considered as a proxy of the reciprocal trust of banks in each other. In a sense, the change in the trend pointed to the erosion of such trust post-crisis. The picture changes by using an entropy-based null-model. Z-scores were used to measure the agreement of the real system with the expectations of the null model, counting the number of standard deviation the observed value is far from the average. The null model used was the Directed Configuration Model (\emph{DCM}), which uses as constraints the in- and out- degree sequences. The trend of the z-scores of the reciprocity relative to the DCM still showed a drop when the crisis hits the network, but it displayed a decreasing trend in the 4 years preceding the arrival of the crisis. In this sense, the system was already experiencing a decreasing phase, in term of trust among banks, few years before the crisis and {\em only the DCM was able to detect it}. 
Similar patterns were observed for other quantities, such as the cyclic motif (three banks involved in a cycling lending pattern) whose presence increases in the period pre-crisis, becoming statistically significant when the crisis hits\cite{squartini2013early}.
Those are the so-called {\it early warning signals}: before a drastic change in the structure of the network, it is still possible to detect smooth changes, which can be highlighted by observing the disagreement between a proper null model and the real system. In the present case, the transition was toward a more fragile structure in the year before the crisis, that reduced the resilience of the system. 

The same framework has been used to analyse the patterns of common asset holdings by financial institutions \cite{gualdi2016statistically}, with the idea that the portfolios overlap not compatible with the null model were carrying the highest riskiness for fire sales liquidation. In order to account for the heterogeneity of financial institutions and owned securities, the null model employed was the Bipartite Configuration Model (\emph{BiCM},~\cite{Saracco2015,Saracco2016b}) that discounts the information carried by the degree sequence of both layers of the bipartite institutions-securities network. The outcome of the analysis was that portfolio similarity was significantly increasing far before the global financial crisis, and peaked at its onset, in a way not compatible with the null model. In other words, the properties of the system were not explainable just by looking at the heterogeneity of portfolios and securities diversification: the system is strongly out of the equilibrium and the observation of significant portfolios overlaps is carrying an extra information~\cite{gualdi2016statistically,Saracco2016b}.

While using a null-model as a benchmark is a standard procedure in order to detect some information on the system, the choice of an entropy-based one is not. Due to the lack of bias by definition~\cite{jaynes1957information}, entropy-based null models are extremely good candidates. In principle, nevertheless, any null model can be used: in this case, the highlighted features of the system just detect the disagreement of reality respect to the model. In some cases, for instance, other methods inspired by statistical physics can be proved to be unbiased~\cite{Tumminello2011, Strona2014, Carstens2015}. In all the aforementioned models, instead of fixing \emph{on average} the constraints of the real system, they are fixed \emph{exactly}. Unfortunately, if the constraints are not informative regarding the nature of the phenomenon under study, the comparison may lead to unsupported conclusions. 

Reference~\cite{Saracco2016b} points out the issues potentially emerging when using a null model in the wrong context. This is the case, for instance, of the projection algorithm presented in~\cite{Tumminello2011} and applied to the ITN. The null model of~\cite{Tumminello2011} applies to bipartite networks where nodes, on one of the layers, are gathered in classes homogeneous by degree. In this sense, the null model preserves the degree sequence of both layers and the degree sequence on one of the layers for each (homogeneous) class of nodes; beside not being easily construable, in the case of the ITN this second kind of information is not crucial for the description of the system. Indeed, the model of~\cite{Tumminello2011} just has too many unjustified constraints and is not able to project any link on any of the two layers. Nevertheless, when applied to the right systems (i.e.\ the ones for which a clear interpretation of the constraints indeed exists), the same null model allows one to gain a deep insight on the considered phenomena~\cite{Tumminello2012,Iori2015,Curme2015a}.

\begin{mdframed}[backgroundcolor=ly]
\medskip

\section*{Box 2: Statistical physics of real-world networks}

The most effective approaches to solve the network reconstruction problem are the ones resting upon the so-called \emph{maximum entropy principle} (MEP) that prescribes to maximize the uncertainty about the system $\mathbf{G}^*$ under analysis while preserving the accessible information. Here, we discuss the choice to model uncertainty via the \emph{Shannon entropy} score function. In this case, the MEP prescribes to maximize

\begin{equation}
S=-\sum_{\mathbf{G}\in\Omega}P(\mathbf{G})\ln P(\mathbf{G})\label{eq:entropy}
\end{equation}
over the ensemble $\Omega$ of graphs with the same number $N$ of nodes and (type of) links of $\mathbf{G}^*$, subject to the normalisation condition $\sum_{\mathbf{G}\in\Omega}P(\mathbf{G})=1$ as well as to a collection of constraints $\mathbf{C}$ representing the information to be preserved \cite{jaynes1957information}. The aforementioned recipe defines the \emph{Exponential Random Graph} (ERG) formalism \cite{park2004statistical,squartini2011analytical,fronczak2014exponential,fronczak2014fluctuation} and ensures that $P(\mathbf{G})$, i.e.\ the occurrence probability of a graph $\mathbf{G}\in\Omega$, is the least biased (i.e.\ maximally non-committal with respect to the properties that are not enforced) probability distribution over the set of allowed configurations of the system under study \cite{jaynes1957information}.

Once the constraints are imposed in a \emph{soft} fashion, by solely requiring that their expected value over the ensemble is determined, i.e.\ $\sum_{\mathbf{G}\in\Omega}P(\mathbf{G})\mathbf{C}(\mathbf{G})=\langle\mathbf{C}\rangle$ , the probability distribution $P(\mathbf{G})$ reads
\begin{equation}
P(\mathbf{G}|\mathbf{\theta})=\frac{e^{-H(\mathbf{G},\mathbf{\theta})}}{Z(\mathbf{\theta})},\quad\forall\:\mathbf{G}\in\Omega\label{eq:prob}
\end{equation}
where $H(\mathbf{G},\mathbf{\theta})=\mathbf{\theta}\cdot\mathbf{C}(\mathbf{G})$ is the \emph{Hamiltonian}, $\mathbf{\theta}$ is the set of Lagrange multipliers enforcing the constraints themselves and $Z(\mathbf{\theta})=\sum_{\mathbf{G}\in\Omega}e^{-H(\mathbf{G},\mathbf{\theta})}$ is the \emph{partition function}. 
Eq.\ \ref{eq:prob} defines the so-called \emph{canonical ensemble}: notice that $P(\mathbf{G}|\mathbf{\theta})$ depends on $\mathbf{G}$ only through the vector of enforced quantities $\mathbf{C}(\mathbf{G})$.

Eq.\ \ref{eq:prob} only specifies the functional form of the probability distribution defining the canonical ensemble while leaving the Lagrange multipliers (numerically) undetermined. A possible strategy is that of drawing the Lagrange multipliers from some \emph{ad-hoc} probability density function, a choice that has been proven to induce archetypal classes of networks (e.g.\ regular graphs, scale-free networks, etc.) \cite{park2004statistical}. However, when the aim is that of fitting a model to the observations, one can invoke the \emph{maximum likelihood principle} (MLP), prescribing to maximise the functional \cite{squartini2011analytical}
\begin{equation}
\mathcal{L}(\mathbf{\theta})=\ln P(\mathbf{G}^*|\mathbf{\theta})=-H(\mathbf{G}^*,\mathbf{\theta})-\ln Z(\mathbf{\theta})\label{eq:likelihood}
\end{equation}
with respect to $\mathbf{\theta}$. Explicitly carrying out the calculations allows to recover the values $\mathbf{\theta}^*$ ensuring that $\langle\mathbf{C}\rangle=\mathbf{C}^*$, implying that the ensemble average of each constraint matches its empirical value $\mathbf{C}^*=\mathbf{C}(\mathbf{G}^*)$.\\

The formalism above has been recently extended in a conditional fashion to provide an estimation of the (purely) weighted structure of networks given any kind of binary information, treated as \emph{prior information} \cite{parisi2020faster}. The score function to optimize now becomes the \emph{conditional Shannon entropy}
\begin{equation}
S(\mathcal{W}|\mathcal{A})=-\sum_{\mathbf{A}\in\mathbb{A}}P(\mathbf{A})\int_{\mathbb{W}_\mathbf{A}}Q(\mathbf{W}|\mathbf{A})\ln Q(\mathbf{W}|\mathbf{A})d\mathbf{W}
\end{equation}
to be maximized with respect to the conditional probability density function $Q(\mathbf{W}|\mathbf{A})$. The normalization constraint now reads $\int_{\mathbb{W}_\mathbf{A}}Q(\mathbf{W}|\mathbf{A})d\mathbf{W}=1$ while the enforcement of a collection of constraints $\mathbf{C}$ can be properly formalized, within the new framework, by writing that $\sum_{\mathbf{A}\in\mathbb{A}}P(\mathbf{A})\int_{\mathbb{W}_\mathbf{A}}Q(\mathbf{W}|\mathbf{A})C(\mathbf{W})d\mathbf{W}=\langle\mathbf{C}\rangle$. Notice that the symbol $\mathbb{W}_\mathbf{A}$ indicates the set of weighted configurations which are compatible with the binary information available \emph{a priori}.

As in the usual ERG framework, a recipe is needed to (numerically) determine the Lagrange multipliers. The estimation of the latter ones can be now carried out by invoking a generalized MLP, whose defining functional becomes
\begin{equation}
\mathcal{G}(\mathbf{\theta})=-H(\mathbf{W}^*,\mathbf{\theta})-\sum_{\mathbf{A}\in\mathbb{A}}P(\mathbf{A})\log Z_{\mathbf{A}}(\mathbf{\theta})
\end{equation}
with the partition function now reading $Z_{\mathbf{A}}(\mathbf{\theta})=\int_{\mathbf{W}_\mathbb{A}}e^{-H(\mathbf{W},\mathbf{\theta})}d\mathbf{W}$.

\end{mdframed}

\section{Conclusions and Perspectives} \label{sec:conclusion}
Financial networks as a discipline has emerged as one of the most successful applications of statistical physics to the realm of social sciences. 
The general approach has been to tackle questions in finance by adapting techniques from fields such as complex networks and statistical physics and by developing novel methods when needed. This has led to the emergence of a new paradigm to understand how risk can propagate, dampened or amplified, across the financial system.
Over the last decade, many concepts, models and metrics developed in the discipline have been adopted both by scholars in the economics profession and by practitioners and policy makers in the financial sector. 
Early policy applications of financial networks were conducted by institutions such as Bank of England\cite{haldane2009rethinking} and the National Bank of Austria\cite{boss2004network}. In the aftermath of the 2008 financial crisis, the policy community and the academia became widely aware that understanding and managing risk in the financial system requires to model it in terms of financial networks. Over the last decade, financial network models have been increasingly used by institutions  
among which the European Central Bank (to assess systemic risk\cite{henry2013macro}), the European Systemic Risk Board (to characterise the derivative market\cite{brunnermeier2013assessing,abad2016shedding} and the network of insurers\cite{alves2015network}), the Office of Financial Research\cite{bookstaber2016looking}, and the Bank of England (to capture feedback mechanisms in the stress test\cite{churm2019four} due to solvency contagion\cite{boe2016stress,boe2017stress} and to funding contagion and overlapping portfolios\cite{boe2017stress}).

The recognition of the importance of network effects has led in the aftermath of the 2008 financial crisis to key conceptual developments in policy. The microprudential regulation (in the finance policy jargon, an approach to regulation focusing at banks individually) has been since then complemented by the macroprudential regulation, which looks at the financial system as a whole (i.e.\ as a network of financial institutions \cite{BCBS2013macro,batiz2016calibrating}) and seeks to limit the impact of financial shocks to the real economy.
The latter approach recognizes that interconnectedness 
(modelled though networks) can have procyclical impact (i.e.\ reinforcing feedbacks) on asset prices, or, in other words, it can amplify risk. Increasingly, stress-tests performed by financial authorities have made use of network models \cite{BCBS2015supervisory}. 

Furthermore, even the Bank of International Settlements (BIS, the institution coordinating banking regulation worldwide) included concepts of financial networks in its metrics to identify systemically important banks\cite{BCBS2013global} and recognised that network effects were responsible for losses on banks’ balance sheets due to valuation adjustment (i.e.\ the recursive repricing of contracts along the financial network).

From the review of the literature the following challenges and avenues for future research stand out. A first challenge comes from the fact that, as described earlier, the financial system is best described as a temporal multiplex networks. The stability properties of the system depend on the interplay of the contagion processes across layers. This is a key topic that deserves further investigation in the future. Indeed different layers may display different contagion processes (e.g.\ dampening or amplifying shocks) and different structures and it is not well understood how to describe the dynamics in a single model. The empirical analysis of such temporal multiplex networks comes with the challenges of handling very large and heterogenous datasets. 

A second challenge for the field comes from trying and modelling the evolution of financial networks over time. Differently from other networks, here nodes have some ability to anticipate the future, including the future structure of the network and try to take advantage of it. Although in reality such ability is much more limited than what models in mainstream economics postulate, it represents a circularity in the model that is difficult to treat with analytical or statistical tools. The study of exogenously given network structures remains however a precondition to understand financial stability. 

A third challenge arises from the growing importance in academia and practice of climate-related financial risk and investments \cite{battiston2017climate}. In this context, risk propagate through long chains of contracts going from physical assets such as fossil-fuel plants or electricity plants scattered around the planet to owners, securities, investors and ultimately, ordinary people savings. There are many open questions both empirical and theoretical that remain to be addressed in this young and dynamic area of the field.

\section*{Acknowledgements}
Diego Garlaschelli acknowledges support from the Dutch Econophysics Foundation (Stichting Econophysics, Leiden, the Netherlands) and the Netherlands Organization for Scientific Research (NWO/OCW).
Guido Caldarelli acknowledges support from the EU project nr.\ 952026-HumanE-AI-Net.

\section*{Author contributions}
All authors contributed equally to this manuscript.

\end{document}